\documentstyle[twocolumn,aps,prl]{revtex}

\input epsf

\begin{document}
\draft

\title{Critical Behaviour in Gravitational Collapse of a Yang-Mills Field}

\author{Matthew W.\ Choptuik}
\address{Center for Relativity, The University of Texas at Austin, Austin,
 TX\ \ 78712-1081}
\author{Tadeusz Chmaj$^{*}$}
\address{Princeton University Observatory, Princeton, NJ\ \ 08544-1001}
\author{Piotr Bizo\'n}
\address{Department of Physics, Jagiellonian University, Cracow, Poland}

\date{\today}

\maketitle

\begin{abstract}
We present results from a numerical study 
of spherically-symmetric collapse of a self-gravitating,
SU(2) gauge field.  Two distinct critical solutions are
observed at the threshold of black hole formation.
In one case the critical solution is discretely self-similar 
and black holes of arbitrarily 
small mass can form.  However, in the other instance the critical
solution is the $n=1$ static Bartnik-Mckinnon sphaleron, and black
hole formation turns on at finite mass.  The transition between these 
two scenarios is characterized by the superposition of both types 
of critical behaviour.
\end{abstract}

\narrowtext

In a recent numerical study of gravitational collapse of a massless scalar
field, a type of critical behaviour was found at the 
threshold of black hole formation~\cite{matt}. More precisely, in the
analysis of spherically-symmetric evolution
of various one-parameter families of initial data describing
imploding scalar waves, it was observed
that there is generically a critical parameter
value, $p=p^{\star}$, which signals the onset of black hole formation. 
In the subcritical regime, $p<p^{\star}$, 
all of the scalar field escapes 
to infinity leaving flat spacetime behind, while for supercritical
evolutions,  $p>p^{\star}$, black holes form
with masses well-fit by a scaling law, 
$M_{BH} \propto (p - p^{\star})^{\gamma}$.
Here, the critical exponent, $\gamma \simeq 0.37$, is universal in the 
sense of being independent of the details of the initial data.
Thus, the transition between 
no-black-hole/black-hole spacetimes may be viewed as a continuous 
phase transition with the black hole mass playing the role
of order parameter.
In the intermediate asymptotic regime 
(i.e. before a solution ``decides'' whether or not to
form a black hole) near-critical
evolutions approach a universal attractor, called the critical solution,
which exhibits discrete self-similarity (echoing).
Using the same basic technique of studying 
families which ``interpolate'' between no-black-hole and black-hole
spacetimes, similar critical behaviour has been observed in 
several other models of gravitational collapse~\cite{grwave1,evans,hhs}.

In this letter we summarize results from numerical study of the evolution
of a self-gravitating non-abelian gauge field modeled by the $SU(2)$
Einstein-Yang-Mills (EYM) equations.  In addition to its intrinsic 
physical interest, we have chosen this model since,
in contrast to all previously studied models, it contains
static solutions which we suspected could 
affect the qualitative picture of critical 
behaviour. Let us recall that these static solutions, discovered by Bartnik 
and Mckinnon (BK)~\cite{bartnikon,joel},
form a countable family $X_n$ ($n \in N$) of spherically-symmetric,
asymptotically flat, regular, but unstable, configurations.

\begin{figure}
\epsfxsize=7cm
\centerline{\epsffile{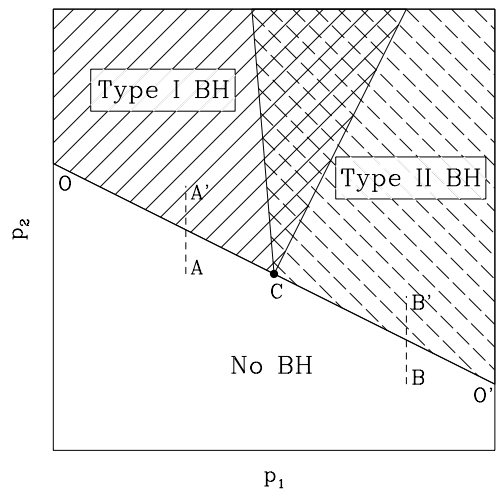}}
\caption{Schematic representation of ``phase-space'' for spherically
symmetric Yang-Mills collapse, showing possible end states of
evolutions from a sufficiently general two-parameter family of
initial conditions. The critical line $OO'$ demarks the threshold
of black hole formation.  An interpolating family such as $AA'$
exhibits Type I behaviour: the critical solution is the static BK
solution $X_1$, and the smallest black
hole formed has finite mass.  Families such as $BB'$ exhibit Type II
behaviour:  the critical solution is discretely self-similar ($\Delta
\approx 0.74$), black hole formation turns on at infinitesimal mass,
and mass-scaling with $\gamma \approx 0.20$ is observed.  At $C$,
the two types of critical behaviour coexist.
}
\label{FIG1}
\end{figure}

Our main new result is the observation that for certain families of
initial data the static
 BK solution $X_1$ plays the role of a critical
solution separating collapse from dispersal. Since in this case there is
a finite gap in the spectrum of black hole masses, we call this
``Type I'' behaviour (in analogy to a first order phase transition),
to distinguish it from ``Type II'' behaviour 
(which we also observe), where black 
hole formation turns on at infinitesimal mass.
As is shown schematically in Figure 1, sufficiently
general two-parameter families of initial data exhibit both types of
critical behaviour.

We consider spherically-symmetric Einstein-Yang-Mills equations with the
gauge group $SU(2)$.
Following~\cite{matt}, we write the general time-dependent,
spherically-symmetric metric as
\begin{eqnarray}
ds^2 &=& -\alpha^2 (r,t) dt^2 + a^2 (r,t) dr^2 + r^2 d \Omega^2.
\label{METRIC}
\end  {eqnarray}
For the YM field we assume the purely magnetic ansatz which in the abelian
gauge means that the field strength has the form~\cite{bartnik}
\begin{equation}
\dot W dt \wedge \Omega + W' dr \wedge \Omega - (1-W^2) \tau_3 d\vartheta
\wedge \sin \vartheta d\varphi ,
\end{equation}
where $\Omega = \tau_1 d\vartheta + \tau_2 \sin \vartheta d\varphi$, $\tau_i$
($i=1,2,3$) are Pauli matrices,  an overdot denotes
$\partial / \partial t$ and a prime denotes
$\partial / \partial r$. Thus the matter content of the model is described
by a single function, $W(r,t)$, which we hereafter refer to as the Yang-Mills
 potential. We note that, as follows from (2), the vacua of the YM field
are given by $W=\pm1$.
  We introduce auxiliary YM variables,
$\Phi \equiv W'$ and
$\Pi \equiv a {\dot W} / \alpha$.
  The dynamics of the EYM model can
then be computed using the following set of equations (see~\cite{bartnik}):
\begin{eqnarray}
{\dot \Phi} &=& \left( \frac{\alpha}{a} \Pi \right) ',  \quad
{\dot \Pi}   =  \left( \frac{\alpha}{a} \Phi \right) ' +
                \frac{\alpha a}{r^2} W \left( 1 - W^2 \right),
\label{PHIDOT}
\\
\frac{a'}{a} &=& \frac{1-a^2}{2r} +
    \frac{1}{r}\left( \Phi^2 + \Pi^2 +
                      \frac{a^2}{2r^2}\left( 1 - W^2)^2 \right)
               \right),
\label{LAPSE}
\\
\frac{\alpha'}{\alpha} &=& \frac{a^2-1}{2r} +
    \frac{1}{r}\left( \Phi^2 + \Pi^2 -
                      \frac{a^2}{2r^2}\left( 1 - W^2)^2 \right)
               \right),
\label{HAM}
\end{eqnarray}
\begin{eqnarray}
W(r,t) &\equiv& \pm 1 + \int^r_0 \Phi({\tilde r},t) d{\tilde r}.
\label{WUPDATE}
\end{eqnarray}

We have solved the initial value problem for many  one-parameter families
of asymptotically flat, regular initial data, some of which are listed 
in Table I. To ensure
regularity at the origin we require that
$W(r,t) \to \pm 1 + O(r^2)$ as $r\to 0$.
The numerical results described below were generated using a 
modified version of the adaptive-mesh algorithm used to perform
the original scalar field calculations~\cite{matt}.  The sensitivity of 
the mesh-refinement algorithm was again very helpful in 
efficiently computing near-critical solutions, but was not 
nearly as crucial as it was for the scalar case.  In fact, we 
have also reproduced most of the following results using a 
uniform-grid code much like the one described in~\cite{zhou}.
 
As stated above, depending on the particular form of initial data used, 
we have found two types of critical behaviour:

\begin{figure}
\epsfxsize=7cm
\centerline{\epsffile{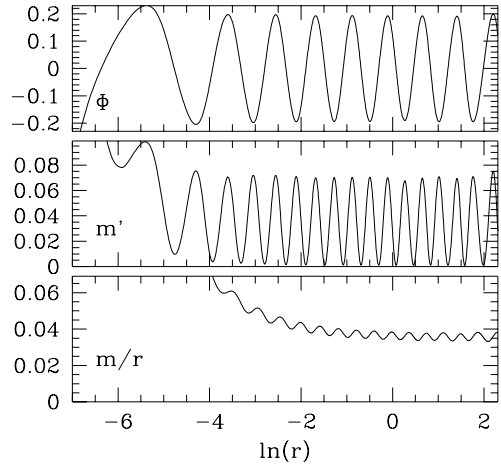}}
\caption{Late time profiles of marginally subcritical Type II collapse
using family~(a) (see Table~1).  Note that the mass aspect,
$m(r,t)$, is defined via $a^2 = (1 - 2m/r)^{-1}$. The large number
of echoes visible here (for fixed $|p-p^\star|/p^\star$), relative
to the scalar case, is a reflection of the relatively small
value of the echoing exponent ($\Delta_{\rm YM} \approx 0.74$ versus
$\Delta_{\rm SF} \approx 3.44$).  However, this is partly offset by
the fact that the mass-scaling exponents for the two models
also differ significantly.  In general, dimensional/scaling
considerations suggest that $\Delta \delta n = -\gamma \delta\pi$
where $\delta n$ and $\delta\pi$ are the changes in echo number, $n$,
and $\pi\equiv \ln|p-p^\star|$, respectively [17].
}
\label{FIG2}
\end{figure}

\begin{figure}
\epsfxsize=7cm
\centerline{\epsffile{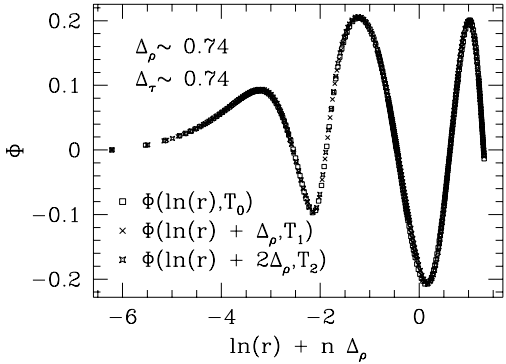}}
\caption{Illustration of scale-periodicity of Type II solution.
This plot shows the superposition of a near-critical
profile of $\Phi$ (at a particular time) with the first two echoes which
subsequently develop.  The central proper time, $T_0$, at which the
earliest profile is monitored is arbitrary; times $T_1$ and $T_2$ and
the rescaling exponent, $\Delta_\rho$, are then chosen to
minimize $\Phi(\ln(r) + n\Delta_\rho,T_n)-\Phi(\ln(r),T_0)$.  An independent
estimate of $\Delta$ is generated by first estimating the critical
time, $T^\star$ for the family, and then computing
$\Delta_\tau \equiv \ln((T^\star - T_n)/(T^\star - T_{n+1}))$.
}
\label{FIG3}
\end{figure}

{\em Type II behaviour.}---In this case we observe a continuous 
no-black-hole/black-hole phase transition, and the overall picture
of criticality is very much analogous to that of scalar field collapse.
For a generic 
Type II family, and in the near-critical, non-linear regime,
we conjecture that the evolution asymptotes to a locally unique
(up to $r, t \to \sigma r, \sigma t$, for arbitrary $\sigma > 0$)
 discretely self-similar solution, with an echoing 
exponent, $\Delta \approx 0.74$.  As with the scalar field 
case, we expect that the precisely critical solution echos
an infinite number of times, exhibits unbounded growth
of curvature near $r=0$, and is singular at the origin at 
some finite value of central proper time, $T^\star$. 
Figure~\ref{FIG2} show profiles of various echoing quantities at 
$T\approx T^\star$ from a family~(a) calculation
(see Table~\ref{TAB1}) with $|s-s^\star|/s^\star \approx 10^{-15}$.  Typical
evidence for scale-periodicity is shown in Figure~\ref{FIG3},
using near-critical data from family~(b) with parameter $a$ varying.
The level of agreement
between the two independently computed estimates of $\Delta$ provides us 
with a rough measure of the accuracy of our computation of the exponent.

\begin{figure}
\epsfxsize=7cm
\centerline{\epsffile{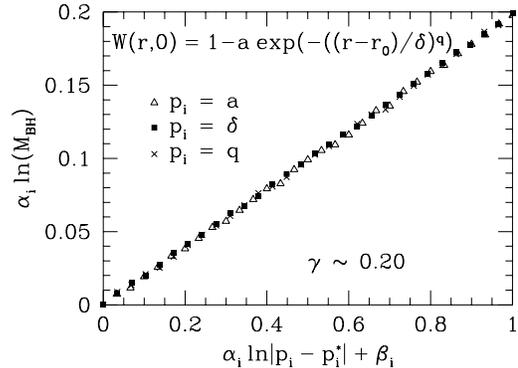}}
\caption{Mass-scaling of Type II solutions.
Each marker type corresponds to a different family of super-critical
computations.  For each family, constants
$\alpha_i$ and $\beta_i$ are chosen to unit-normalize the $x$-range
and place the first data-point (smallest black hole) at the origin.
For all plots, the least squares fit for the slope, $\gamma$,
is 0.20, with an estimated uncertainty of a few percent.  In addition,
for all families, the unnormalized $\pi$ range was 18; thus in each
case, the black hole mass spans a factor of $e^{18\gamma} \approx 37$.
}
\label{FIG4}
\end{figure}

As expected, Type II families also exhibit mass-scaling in the 
super-critical regime: 
$M_{BH} \propto (p - p^{\star})^{\gamma}$.
Typical results are shown in Figure~\ref{FIG4} and 
we estimate that the value $\gamma\approx0.20$ is accurate to a few 
percent.  We note in passing that this result is another piece of 
the growing body of evidence which has shown that $\gamma$ is {\em not\/}
constant across all collapse models~\cite{maison,hhs,steve}.
As argued in~\cite{koike},
mass-scaling and universality (initial-data-independence), 
strongly suggest that the stable manifold of a Type II solution, such 
as the one described here, is of codimension one.  
This picture, which predicts that $\gamma$ is the
reciprocal of the Lyapounov exponent of a single growing mode
associated with a critical solution,
has now been validated for at least two distinct matter models with
{\em continuously\/} self-similar critical
solutions~\cite{koike,ehh,hhs,he1,he2}.
Perturbative treatment of critical solutions with discrete symmetry is 
more involved, although considerable 
progress has been made for the scalar case~\cite{gundlach}. Here 
we only remark that any 
techniques which work for the scalar field should be largely 
applicable to this solution.

\begin{figure}
\epsfxsize=7cm
\centerline{\epsffile{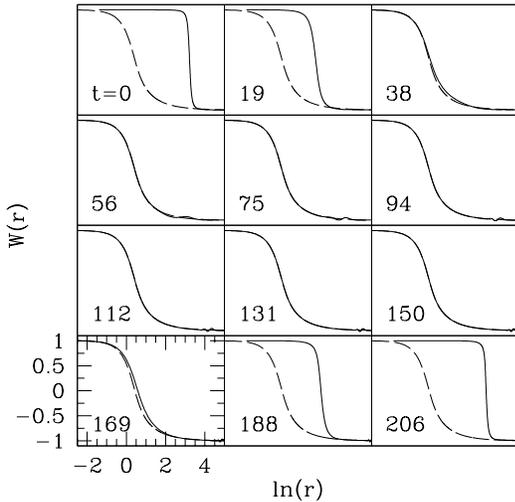}}
\caption{Marginally subcritical Type I evolution.  Here we plot the dynamical
evolution of $W(r,t)$ (solid line) and superimpose the static
BM configuration $X_1$ (dashed line).
 Initially, the evolution (family (c),
$|\delta-\delta^\star|/\delta^\star \approx 10^{-15}$) is nearly
linear and almost purely ingoing.  When the pulse arrives at
the center, it sheds off YM radiation, approaches $X_1$ and stays near it
 for some time, and then disperses to infinity.
}
\label{FIG5}
\end{figure}

{\em Type I behaviour.}---As stated previously, solutions in a Type I 
interpolating family asymptote to the {\em static}  BK solution $X_1$ as 
$p\to p^\star$. 
This behaviour appears generically for kink-type initial profiles
of the YM potential $W$, such as family (c) in Table I. 
In this case it has already been established that the each of 
the BK solutions, $X_n$,
has exactly $n$ unstable modes (within the ansatz (2)); hence 
we are certain that the stable manifold of the critical solution 
has codimension one.
Initial data with small $|p-p^\star|$ results in an 
evolution which approaches
$X_1$ and  stays in its vicinity for central proper time,
 $T \approx -\lambda \ln|p-p^{\star}|$.
The configuration then 
either disperses to infinity ($p<p^{\star}$) or collapses to
a black hole with finite mass ($p>p^{\star}$). 
Typical results from a marginally subcritical Type I evolution are shown in 
Figure~\ref{FIG5}.

Our calculations are somewhat complementary to 
those performed by Zhou and Straumann~\cite{zhou}. Those authors generally 
used initial conditions describing small deviations from the static
solution $X_1$, and studied the subsequent evolution to verify the 
prediction of perturbative instability of $X_1$~\cite{straumann}.  Here, by
construction, we {\em generate} the solution $X_1$ as the boundary between 
collapse and dispersal, and thus immediately verify its instability.
In this case, the reciprocal Lyapounov exponent of the single unstable mode
yields the
characteristic time scale, $\lambda$, for the decay of $X_1$.
Using central-proper-time normalization, the value computed 
from perturbation theory~\cite{straumann} is 
$\lambda = 0.5519\ldots$.  

We can measure this exponent quite 
directly from our simulations by computing the variation of the 
lifetime of near-critical
solutions with respect to variations in $\pi \equiv \ln|p-p^\star|$.
Specifically, defining $T_r(\pi)$ to be the central proper time 
at which the zero-crossing of $W(r,t)$ reaches radius $r$ as it 
propagates outwards, we expect $-dT_r/d\pi \to \lambda$ as $\pi\to-\infty$
and for sufficiently large $r$.
Some numerical regularization is provided 
by monitoring $T_r$ at several discrete radii $r_i, i = 1 ... n$ and 
computing the averaged quantity 
${\bar T}_r(\pi) \equiv n^{-1}\sum T_{r_i}(\pi)$. When this is done 
for family~(c) with $r_1=400,r_n=475,n=16$, we find 
$0.5525 < -d{\bar T}_r/d\pi < 0.5520$ for $-19 < \pi <-10$.  In addition,
we can get a good estimate of the unstable {\em eigenmode\/} by studying
near-critical departures from the static solution. 

As noted in the introduction, Type I behaviour is clearly characterized by a 
gap in the black hole mass spectrum at threshold.  We conjecture that the 
mass-gap is 
universal, and observe that our calculations suggest that it is 
very close (1\% or less) to the total mass of the static solution
$(m_1 = 0.828640 \ldots)$.

{\em Type I and II coexistence.}---We can only briefly describe what 
is one of the more interesting
features of critical behaviour in the EYM model: the fact 
that for certain two parameter families---such as (a) and (d) in 
Table~\ref{TAB1}---there exists a critical 
line in parameter space (see Fig~\ref{FIG1}) which interpolates 
between Type I and Type II behaviour.  The results we have obtained 
lead us to conjecture that the transition point ($C$ in Fig~\ref{FIG1})
represents coexistence of the two distinct critical solutions described
above.  In other words, near $C$, we see echoing occurring in 
context of the static, $n=1$ background.  Thus we can have
have arbitrarily small black hole formation within a configuration
which itself is arbitrarily close to forming a finite-mass black hole.
We end by noting that the super-critical regime in this model,
especially in the cross-hatched overlap region sketched in Fig~\ref{FIG1} is 
still very much {\em terra incognita}, although we have preliminary 
evidence that further phenomenological richness lurks there. 
In particular, we have observed there an intriguing discontinuity in
the spectrum of black-hole masses, which suggests the existence of
another type of critical behaviour in the formation of 
black holes~\cite{chmaj}.

{\em Acknowledgments.} MWC would like to thank 
Peter Forg\'acs for helpful discussions. 
We acknowledge the hospitality of the Erwin Schr\"odinger Institute
in Vienna, where this work was initiated.
The research of MWC was supported in part by NSF PHY9310083, PHY9318152 and 
Metacenter grant MCA94P015 and by a Cray Research Grant to R.~Matzner.
The research of TC and PB was supported in part by the KBN grant
PB750/P3/94/06. 


\newpage
\widetext

\begin{table}
\caption{Initial data families used in the calculations reported here.
Listed are the family label, the form of the initial profile, $W(r,0)$, 
family parameters, $p_i$, nature of $W$'s initial time derivative, and 
which behaviour is found by generating interpolating families.
Initial data for family (a) is time-symmetric (${\dot W(r,0)} = 0$);
for all other families the initial configurations are almost purely
ingoing.  Finally, for family (a), the constants $a$ and $b$ are chosen 
so that $W(0,0)=1$ and $W'(0,0)=0$. 
}

\begin{tabular}{ddddd}
Family & $W(r,0)$ & $p_i$ & ${\dot W(r,0)}$ & Behaviour \\
\tableline
(a) &   $(1 + a\, (1 + b r/s) \exp(-2 (r/s)^2)) \tanh((x-r)/s)$
    &   $x, s$           & 0                & I, II \\
(b) &   $1 + a \exp(-((r-20)/\delta)^q)$
    &   $a, \delta, q$   & $W'(r,0)$        & II    \\
(c) &   $(1 - (r/\delta)^2) / ((1 - (r/\delta)^2)^2 + 4r^2)^{1/2}$
    &   $\delta$         & $W'(r,0)$        & I     \\
(d) &   $-1 + 2 a \exp(-((r - 17)/4)^q)$
    &   $q,  a$          & $W'(r,0)$        & I, II  \\
\label{TAB1}
\end{tabular}
\end{table}

\begin{references}

\bibitem[*]{}
Kosciuszko Foundation Fellow. On leave of absence from
N. Copernicus Astronomical Center, Cracow, Poland.

\bibitem{matt} M.\ W. Choptuik,
 Phys. Rev. Lett. {\bf 70}, 9-12 (1993).

\bibitem {grwave1} A.M. Abrahams and C.R. Evans,
 Phys. Rev. Lett. {\bf70}, 2980-2983 (1993).

\bibitem{evans} C.R. Evans, and J.S. Coleman,
 Phys. Rev. Lett. {\bf72}, 1782-1785 (1994).

\bibitem{hhs} R.~S.Hamade, J.~H.~Horne, and J.~M.~Stewart,
  LANL preprint gr-qc/9511024 (1995).

\bibitem{bartnikon} R. Bartnik and J. Mckinnon, 
Phys. Rev. Lett. {\bf 61}, 141 (1988).

\bibitem{joel} J. A. Smoller and A. Wasserman,
 Commun. Math. Phys. {\bf 151}, 303 (1993).

\bibitem{bartnik} R.~Bartnik, Proceedings of the Third Hungarian
Relativity Workshop, ed. Z. Perjes (Budapest, World Scientific, 1990).

\bibitem{zhou} Z.-H.~Zhou and N.~Straumann, 
Nucl. Phys. {\bf B360}, 180 (1991);
Z.~Zhou, Helv. Phys. Acta. {\bf 65}, 767-819 (1992).

\bibitem{maison} D.~Maison, 
Phys. Lett. {\bf B366}, 82-84, (1996). 

\bibitem{steve} M~W.~Choptuik and S.~L. Liebling,
  in preparation, (1996).

\bibitem{koike} T.~Koike, T.~Hara and S.~Adachi, 
 Phys. Rev. Lett. {\bf74}, 5170-5173 (1995).

\bibitem{ehh} D.~Eardley, E.~Hirschmann and J.~Horne,
  Phys. Rev.  {\bf D52}, 5397-5401 (1995).
        
\bibitem{he1} E.~Hirschmann and D.~Eardley,
  Phys. Rev. {\bf D52}, 5850-5856 (1995).

\bibitem{he2} E.~W.~Hirschmann and D.~M.~Eardley,
  LANL preprint gr-qc/9511052 (1995).

\bibitem{gundlach} C.~Gundlach,
  Phys. Rev. Lett. {\bf 75}, 3214-3217 (1995).

\bibitem{straumann} N.~Straumann and Z.-H.~Zhou, 
Phys. Lett. {\bf B243},
33 (1990).

\bibitem{evansps} C.~Evans, personal communication, (1993)

\bibitem{chmaj} P.~Bizo\'n{}, T.~Chmaj and Z.~Tabor in preparation, (1996).

\end{references}
\end{document}